\documentclass[preprint]{aastex}
\usepackage{graphicx, natbib,epsfig,amsmath,amssymb, mathrsfs, txfonts}
\usepackage{epstopdf}
\begin{document}
\title{Band-Limited Coronagraphs using a Halftone-dot process:\\ II. Advances and laboratory results for arbitrary telescope apertures}

\author{P. Martinez\altaffilmark{1}, C. Dorrer\altaffilmark{2}, and M. Kasper\altaffilmark{3}}

\altaffiltext{1}{UJF-Grenoble 1 / CNRS-INSU, Institut de Plan\'{e}tologie et d'Astrophysique de Grenoble (IPAG) UMR 5274, Grenoble, F-38041, France }
\altaffiltext{2}{Aktiwave, 241 Ashley drive, Rochester, NY, 14620-USA}
\altaffiltext{3}{European Southern Observatory, Karl-Schwarzschild-Strasse 2, D-85748, Garching, Germany}

\begin{abstract}
The band-limited coronagraph is a nearly ideal concept that theoretically enables perfect cancellation of all the light of an on-axis source. 
Over the past years, several prototypes have been developed and tested in the laboratory, and more emphasis is now on developing optimal technologies that can efficiently 
deliver the expected high-contrast levels of such a concept. 
Following the development of an early near-IR demonstrator, we present and discuss the results of a second-generation prototype using Halftone-dot technology. We report improvement in the accuracy of the control of the local transmission of the manufactured prototype, which was measured to be less than 1$\%$. 
This advanced $H$-band band-limited device demonstrated excellent contrast levels in the laboratory, down to $\sim10^{-6}$ at farther angular separations than 3$\lambda/D$ over 24$\%$ spectral bandwidth. These performances outperform the ones of our former prototype by more than an order of magnitude and confirm the maturity of the manufacturing process. 
Current and next generation high-contrast instruments can directly benefit from such capabilities. In this context, we experimentally examine the ability of the band-limited coronagraph to withstand various complex telescope apertures.   
\end{abstract}

\keywords{instrumentation: high angular resolution --- techniques: high angular resolution}
 
\section{Introduction}
A large variety of astrophysical topics such as low-mass companions, or circumstellar disks, have driven the next generation of high-contrast instruments such as SPHERE and GPI \citep{2008SPIE.7014E..41B,2006SPIE.6272E..18M}. Typically, detection and spectroscopic characterization of relatively young objects, giant planets, and brown dwarfs are possible around nearby stars.  With the worldwide emergence of Extremely Large Telescope (ELT) projects, older and fainter companions will become accessible. These future ELTs will improve the sensitivity of exoplanet searches toward lower masses and closer angular distances, ideally down to rocky planets.

The band-limited coronagraph \citep[BLC,][]{2002ApJ...570..900K} arises as a relevant concept for the next generation of ground-based observatories since it has the advantage of being less sensitive to the primary mirror segmentation, unavoidable with ELTs, than to other concepts \citep{SIVA05}. The BLC has emerged as a pertinent candidate when compared to a large panel of coronagraphic concepts by means of complex simulations \citep{Corono}. BLCs provide achromatic behavior, which is very important, as smooth and flat chromaticity response of the coronagraph is essential for exoplanet detection, as broadband observations are required for spectroscopy. The BLC is therefore an apposite candidate for any high-contrast instruments on ELTs. The BLC is also considered as a potential second-generation concept for SPHERE. 

However, as for all concepts that theoretically provide perfect cancellation of on-axis starlight, the BLC faces severe construction tolerances.
During the past years several prototypes have been developed  \citep[e.g.][]{Debes04b, Trauger, Creep06, Trauger2, 2008SPIE.7010E.107M} for visible wavelength application, while more recently we examine the use of a halftone-dot technology, namely, the microdot technique, to produce a BLC \citep[][]{2009ApJ...705.1637M}. 
This process allowed us to develop and test the first near-IR BLC demonstrator (hereafter, Proto 1). 
These microdot masks consisting on opaque square pixels (called dots) distributed to reproduce the continuous transmission of a filter have several advantages:
relative ease of manufacture, achromaticity, reproducibility, and the ability to generate continuous transmission ranges without introducing wavefront errors. In addition, the technique is adapted to the manufacture of a BLC regardless of the wavelength range of application (visible, near-IR, IR) as long as the dot size does not fall into a regime where it is comparable to or smaller than the size of the imaging wavelength (sub-wavelength grating situation). 

In \citet[][]{2009ApJ...705.1637M} -- hereafter Paper 1 --,  the principle, properties, and design guidelines for such a mask construction were provided, and laboratory results obtained in the near-infrared were presented.
In this paper, we report improvement in the accuracy of the control of the local transmission of the manufactured microdot prototype with a second-generation prototype (hereafter, Proto 2), and demonstrate improved performance. 
While Proto 1 was tested with a clear aperture (off-axis telescope), the characterization of Proto 2 is extended to the case of present-day (Very Large Telescope) and future telescope (European Extremely Large Telescope) designs. In addition, the response of the BLC to the pupil-stop shape is discussed through the use of a large panel of designs delivering various throughputs. 

In Section 2 we merely recall the theory of the BLC, while in Section 3 we present the halftone-dot technology and discussed the improvement made with our recent $H$-band device. Section 4 is dedicated to the experimental evaluation of the prototype performance. Finally, in Section 5 we draw conclusions.

\section{The Band-Limited Coronagraph}
The BLC was proposed by \citet{2002ApJ...570..900K} to both block light and manage diffraction effects caused by removal of the light.
Since the theory is detailed in several papers (e.g.,  \citet{2002ApJ...570..900K}), we solely provide here a brief overview. 

\noindent We consider the following BLC amplitude mask function:
\begin{equation}
M(r) = N \left( 1 - sinc\left(\frac{\epsilon r D}{\lambda}\right) \right)
\label{function}
\end{equation}

\noindent where $\lambda$ is the wavelength of the application, $r$ is the radial coordinate in the image plane, D is the telescope primary diameter, $\epsilon$ is the bandwidth that rules the inner-working angle of the coronagraph (IWA hereafter), and finally N is a constant of normalization insuring that $0 \leqq M(r) \leqq 1$. 
We can now express its Fourier transform denoted as $\mathscr{M}(u)$:
\begin{equation}
\mathscr{M}(u) = N \left( \delta \left( \frac{u \lambda}{D}\right) - \frac{ \lambda}{\epsilon D} \times \Pi \left(\frac{u \lambda}{\epsilon D} \right)  \right)
\label{function}
\end{equation}
\noindent where $u$ stands for the radial coordinate of the pupil plane, $\delta$ the Dirac-function, and $\Pi$ the top-hat function.

The power of a BLC comes from the properties that  $\mathscr{M}(u)$ is equal to zero everywhere but $\vert u \vert < \frac{\epsilon D}{2 \lambda}$, i.e. the power spectrum of $M(r)$ has power in a limited domain of frequencies (i.e., a band-limited image mask):
\begin{equation}
\mathscr{M}(u) = 0, \hspace{0.2cm}  \mid u \mid < \frac{\epsilon D}{2 \lambda}
\label{condition}
\end{equation}
As a result, the convolution product of $\mathscr{M}(u)$ by the pupil aperture confines the diffracted light in the vicinity of the pupil edges (by diffracting all the light from an on-axis source to angles within $\frac{\epsilon D}{2 \lambda}$ of the edges of the pupil), which can be completely removed with an adequate undersized pupil-stop. 
A BLC has full discovery space, intrinsic achromatic properties, and an adaptable IWA.

\section{Halftone-dot technology}
\begin{figure*}[!ht]
\centering
\begin{center}
\includegraphics[width=10.0cm]{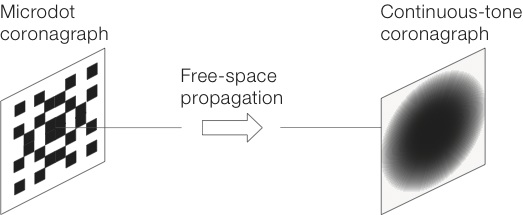}
\end{center}
\caption{Principle of the halftone-dot technology for coronagraphy} 
\label{principle}
\end{figure*}

Controlling the amplitude of light is crucial to many scientific applications, such as those in imaging systems, astronomical instruments, or laser physics. 
In the context of a coronagraph, because it interacts with focused light, such devices experience tighter manufacturing tolerances and issues. 
The halftone-dot technology consists of the process of presenting a continuous image through use of dots (the schematic principle is presented in Fig. \ref{principle}); it originally took its origin in the printing industry. 
With this process, customized filters with spatially varying transmissions are produced using a binary array of metal pixels that offers excellent control of the local transmission. 
This process represents a suitable alternative to other technologies employed for the manufacturing of BLC masks: (1) a grayscale pattern written with a high-energy beam sensitive glass using $e$-beam lithography, (2) a notch filter pattern written with a thick Chromium layer on a substrate, dry etched with high density decoupled plasma, (3/) a grayscale pattern manufactured with vacuum-deposited metals and dielectrics. 
The halftone-dot technology has been selected for the manufacturing of the \textit{James Webb Space Telescope} (JWST)/NIRCam coronagraphs \citep{Krist09}, and for both the SPHERE and GPI coronagraphs (apodizer masks for the apodized-pupil Lyot coronagraph).

\subsection{Principle of a Microdot BLC Mask}
A BLC manufactured in microdots consists in an array of dots (i.e., pixels) that are either opaque or transparent (Fig. \ref{principle}). It is fabricated by lithography of a light-blocking metal layer deposited on a transparent glass substrate. 
To best approximate the selected BLC function, the dot distribution is regulated by a Floyd-Steinberg dithering algorithm \citep{floyd}, based on the error diffusion principle, which allows the accurate generation of grey levels and rapidly varying shaping functions.  This specific algorithm is presented and discussed for printing technique applications \citep{Ulichney, Ulichney2}, for laser beam shaping \citep{2007JOSAB..24.1268D}, and for coronagraphy in the context of the manufacture of pupil-apodizers  \citep{microdots1, microdots2, microdots3}. We note that another algorithm (emphasizing randomize dot distribution) has been successfully used to produce BLC in microdots for the JWST/NIRCam \citep{Krist09}. 

Briefly, we detailed the principle of the error-diffusion algorithm. The error-diffusion algorithm is used to calculate the density distribution that most effectively reproduces the required field transmission; 
the algorithm chooses the transmission of a given pixel of the pattern (either 0$\%$ or 100$\%$) by comparing the transmission required at this location to a 50$\%$ threshold, i.e., the transmission is set to zero if the required transmission is smaller, or to one otherwise. The induced transmission error is diffused into adjacent pixels that have not yet been processed by biasing the transmission required at the corresponding locations. This locally cancels the error of the binary optics introduced by the process of writing the required transmission (in gray levels) into binary values. 
The advantage of the error diffusion algorithm is that the introduced noise is blue, i.e., the noise spectral density is only significant at high spatial frequencies, whereas randomized based dot distribution algorithms introduce white noise. Then fabrication errors enter the game, and multiple variations of the design are required to accommodate best fabrication errors.

In Paper 1, we showed that although total starlight cancellation cannot be achieved with a microdot BLC, the technology does not preclude deep high contrast levels. This impossibility of delivering perfect starlight removal originates from the difference in frequency contents between a microdot BLC and a theoretical continuous mask. The function of the microdot BLC is indeed sampled by the dots, which forces its Fourier transform $\mathscr{M}(u)$ to be periodic, and the algorithm used to distribute dots introduce high-frequency noise. Thus, Eq. \ref{condition} is no longer satisfied. 
\begin{figure*}[!ht]
\begin{center}
\includegraphics[width=5.7cm]{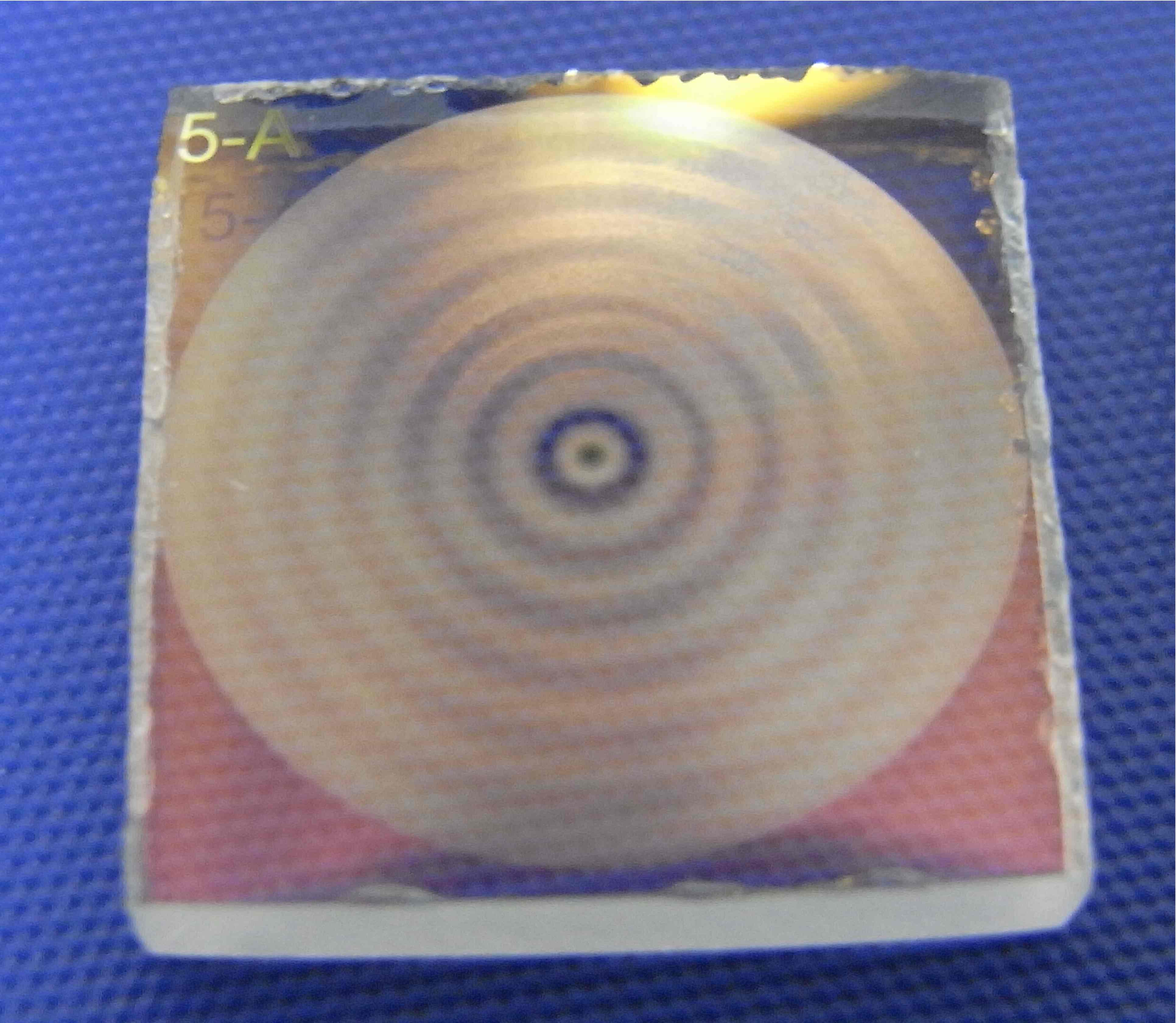}
\includegraphics[width=5cm]{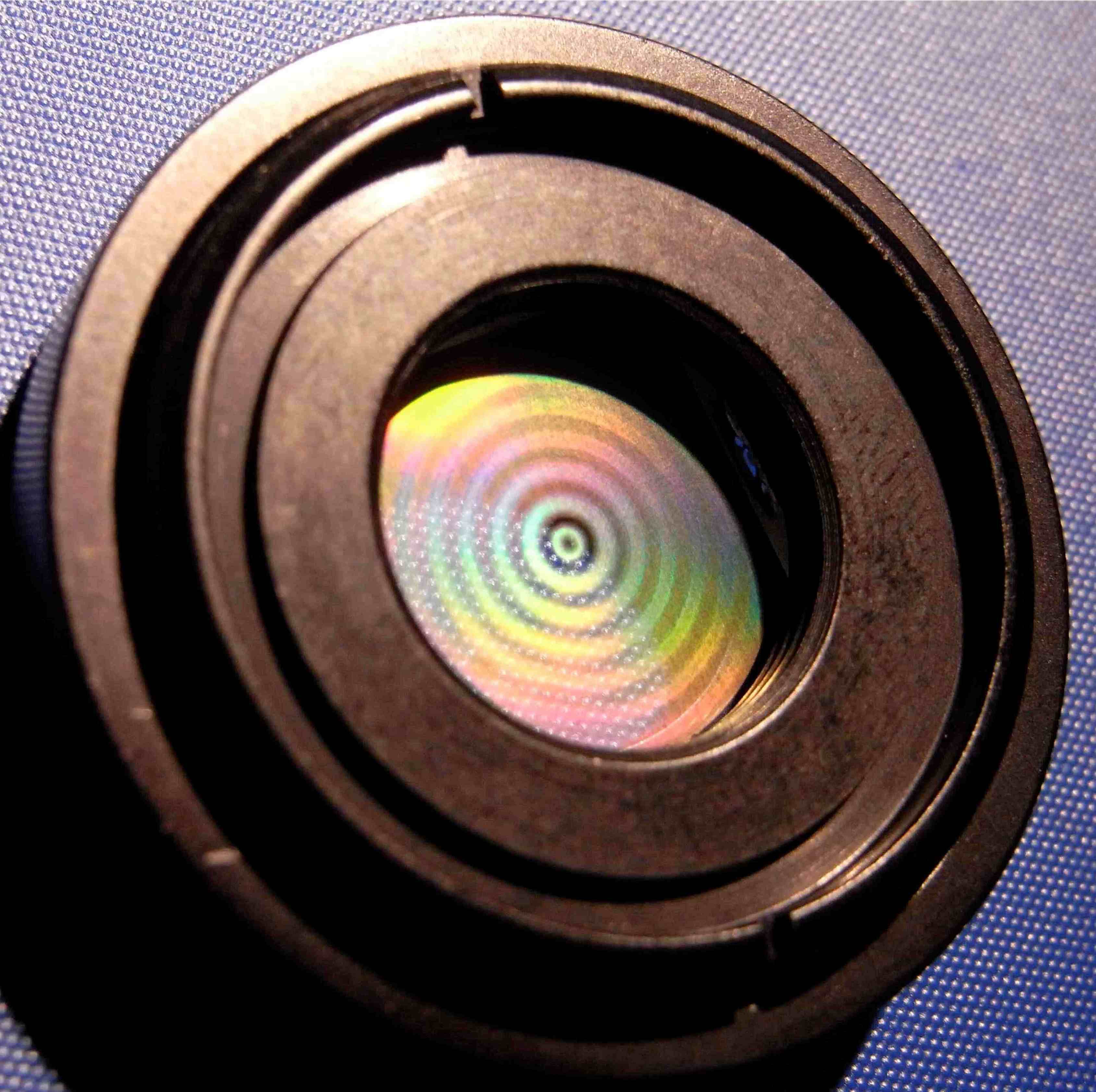}
\end{center}
\caption{Picture of the BL5 prototype (left), and in its integration mount (right).} 
\label{view}
\end{figure*}  
However, contrast levels enabled in such a situation are a function of pixel size, since this allows finer control of the local transmission and is formally equivalent to a sampling problem.
A simple metric was derived in Paper 1 to define the sampling of the BLC function as the ratio of the Airy unit by the dot size ($p$).
This metrics is defined at the shortest wavelength for which the BLC mask is designed to operate, and can be expressed as:
\begin{equation}
s = \frac{F_{\#} \times \lambda_{min}}{p}
\label{E2}
\end{equation}
\noindent where $F_{\#}$ is the $f$-number on the mask and $\lambda_{min}$ the shortest wavelength of the application (i.e., insuring that $s$ remains greater than the specification for all wavelengths). 
Using simulation maps of BLCs designed in microdots (assuming specified dot spatial distribution), we analyzed how the dot size affects the coronagraphic performance with respect to continuous idealistic masks.
We identified two sampling configurations suitable for near-IR operation: $s=16$ and $s=8$. \\
In addition, on the basis of simulations, we pointed out that the interest of the microdot technique in the light of contrast and inner-working angle requirements is not dependent of the BLC function bandwidth ($\epsilon$, which rules the IWA) assuming IWA $\geq$ 3$\lambda/D$, which falls in agreement with EPICS (20-30 mas in $H$-band) and SPHERE (0.1 arcsec in $H$-band) standard requirements.

In this context we developed Proto 1 in two versions, with $s=16$ and $s=8$ configurations. Both have demonstrated impressive and similar performance in the laboratory. 
\begin{figure*}[!ht]
\centering
\begin{center}
\includegraphics[width=8.2cm]{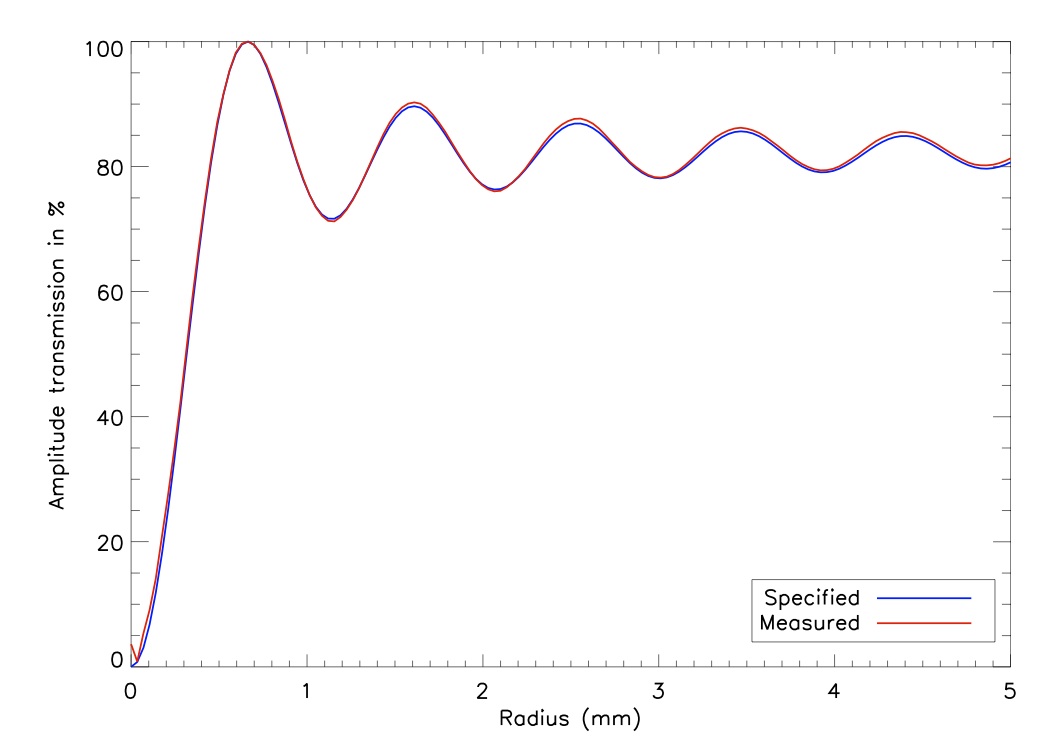}
\includegraphics[width=8.2cm]{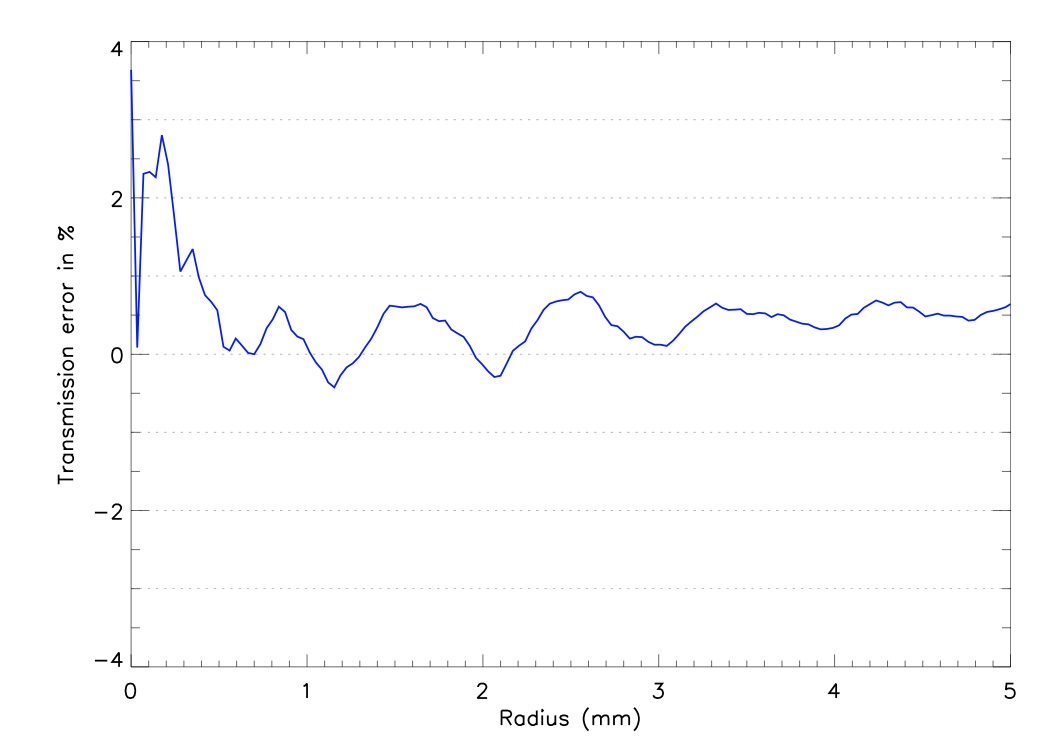}
\end{center}
\caption{Measured profiles of BL5 from the center to the edges of the pattern and their comparison to specifications (left). The corresponding transmission error is as function of the radius (right).} 
\label{res0}
\end{figure*}  
\subsection{Current device}
Proto 2 is based on the same amplitude mask function as Proto 1, being expressed in Equation \ref{function} ($\epsilon=0.17$, defining an inner-working angle of 5$\lambda/D$), and the following microdot parameter: $s=16$ ($p=$5 $\mu$m, $\lambda_{min}$ =1.64 $\mu$m, and $F_{\#}=48.8$).
The BLC amplitude mask function was originally proposed by \citet{2002ApJ...570..900K}.

Like Proto 1, Proto 2 was manufactured by Precision Optical imaging (Rochester, New York),  was designed for the $H$-band, and was fabricated using wet-etch contact lithography of an Aluminum layer ($OD=8+$, $e=2000\dot{\mathrm{A}}$) deposited on a BK7 substrate ($\lambda/10$ peak-to-valley, 0.5 inch diameter). Antireflection coating has been applied for each face ($R<0.5\%$).
Such OD is required to guarantee against leakage greater than the intrinsic limitation of the microdot technique (s = 16, see Paper 1).
A set of pictures of BL5 is shown in Figure \ref{view} before and after integration of the mask onto its mechanical support.
The size of the dots has been confirmed by a microscopic measurement.
Its profile was measured at 1.0$\mu$m, and the spatially resolved transmission has been obtained after background subtraction and flat field normalization.  
Proto 1 profile accuracy was of about $5\%$ from the specification, where the error was mostly localized in the outer part of the mask (high-transmission part), while the center part (for low-transmissions) was highly accurate (Paper 1). The error took its origin in a calibration issue of the process, which has been corrected through several runs to increase the accuracy. Multiple variations of the design on the BL5 mask have been realized to accommodate best fabrication errors. 
\begin{figure}[!ht]
\begin{center}
\includegraphics[width=3.9cm]{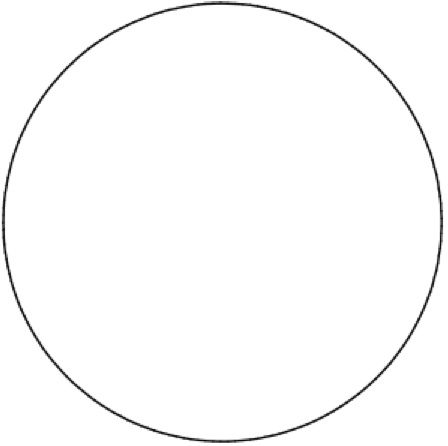}
\includegraphics[width=3.9cm]{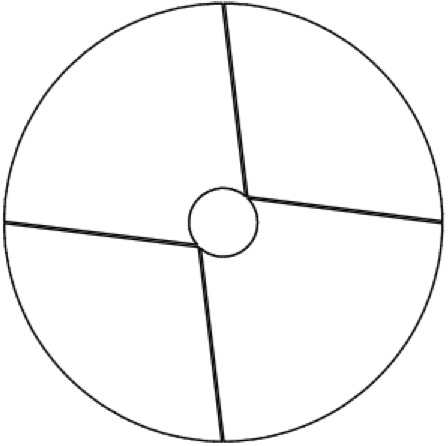}
\includegraphics[width=3.9cm]{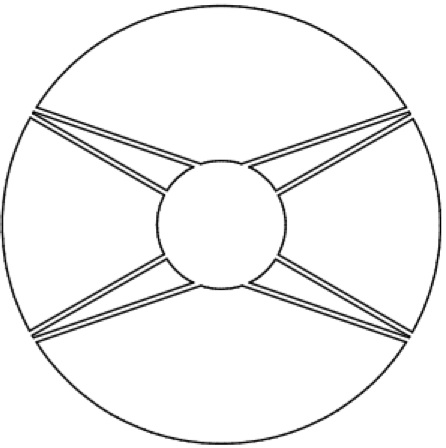} \\
\includegraphics[width=4.cm]{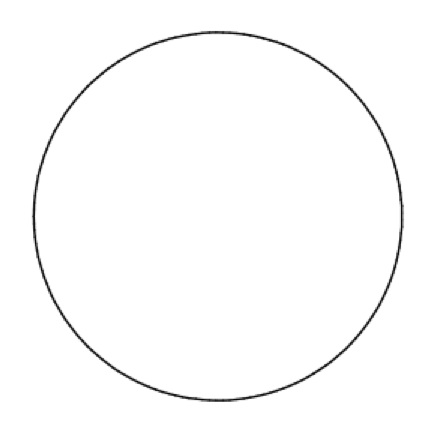}
\includegraphics[width=3.9cm]{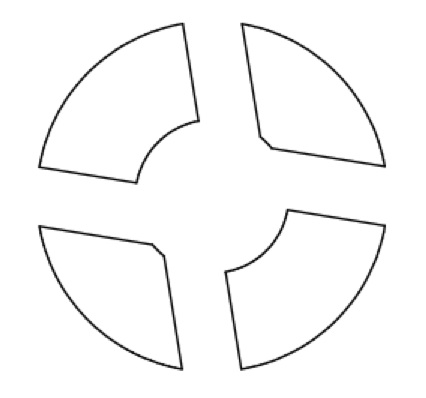}
\includegraphics[width=4cm]{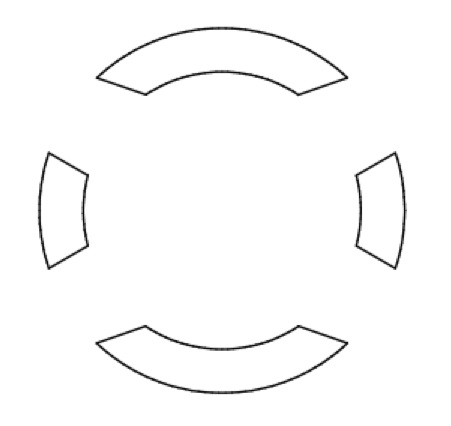}
\end{center}
\caption{Top row: entrance pupil apertures available for the experiment (from the left to the right: clear pupil, VLT pupil, and E-ELT pupil). Bottom row: corresponding BLC pupil-stops.} 
\label{pupil}
\end{figure}  
The profile accuracy of Proto 2 stands below 1$\%$ (Fig. \ref{res0}) everywhere except in the small and central area of the pattern (from the center to 0.4 mm radius) where the error rises to a maximum of 3.6$\%$ at the center and otherwise remains below 3$\%$. 
It is believed that the discrepancy at the center of the BL5 is a measurement bias rather than a real error on the profile. The reason is threefold:
(1) The error terms on the design behave according to the error diffusion algorithm, which is not compatible with an increase of the error around the design transmission close to zero, while fabrication errors occur regardless from the position on the mask. (2) Although a proper measurement of the OD has not been carried out (specification is OD = 8+), the laboratory results obtained with BL5 and presented in the next section guarantee that at least OD = 4 (peak rejection is of about $\sim10^{-4}$). This level of OD limits the transmission leakage at the center of the part;
(3) Because the profile presented in Figure \ref{res0} corresponds to a radial average, the characteristics of the optical setup used for the characterization of the part, the centering and alignment of the BL5 during the measurement does impact its reliability at very low transmission. For instance, as the sampling in the camera plane is finite to approximately 30 $\mu$m, the centering of the part is an important issue.  
Because the intensity is roughly parabolic at the center, with an offset of the center, the radial average then corresponds to a sum of parabola with the same curvature but a slightly different origin. 
This leads to a parabola with the same curvature, but a non-zero origin. This is observed in the data, as there is a non-zero slope at the center of the radial intensity, whereas the slope would be expected to be equal to zero. Since the intensity at the origin is used to remove the background, this leads to a higher than expected value of the measured radial intensity away from the origin, hence the observed error for small transmissions. 

\section{Laboratory evaluation and results}
\subsection{Optical Setup}
The optical setup used to characterize the BL5 prototype is the same as the one presented in Paper 1. Briefly, the setup is a near-infrared coronagraphic transmission bench developed at ESO. All the optics are set on a table with air suspension in a dark room and are fully covered with protection panels forming a nearly closed box preventing turbulence from the room. 
The entrance aperture is 3 mm diameter ($\Phi$), and reimaging optics are made with IR achromatic doublets. The pupil patterns that can be used on the bench are shown in Figure \ref{pupil}.
The IR camera (the Infrared Test Camera) uses a HAWAII $1k\times1k$ detector. The experiment was done in $H$-band using either a narrow filter ($\Delta \lambda/\lambda =1.4\%$) or a broadband filter ($\Delta \lambda/\lambda = 24\%$). 
The Strehl ratio of the bench was evaluated to $\sim92\%$.

\begin{figure*}[!ht]
\begin{center}
\includegraphics[width=8.2cm]{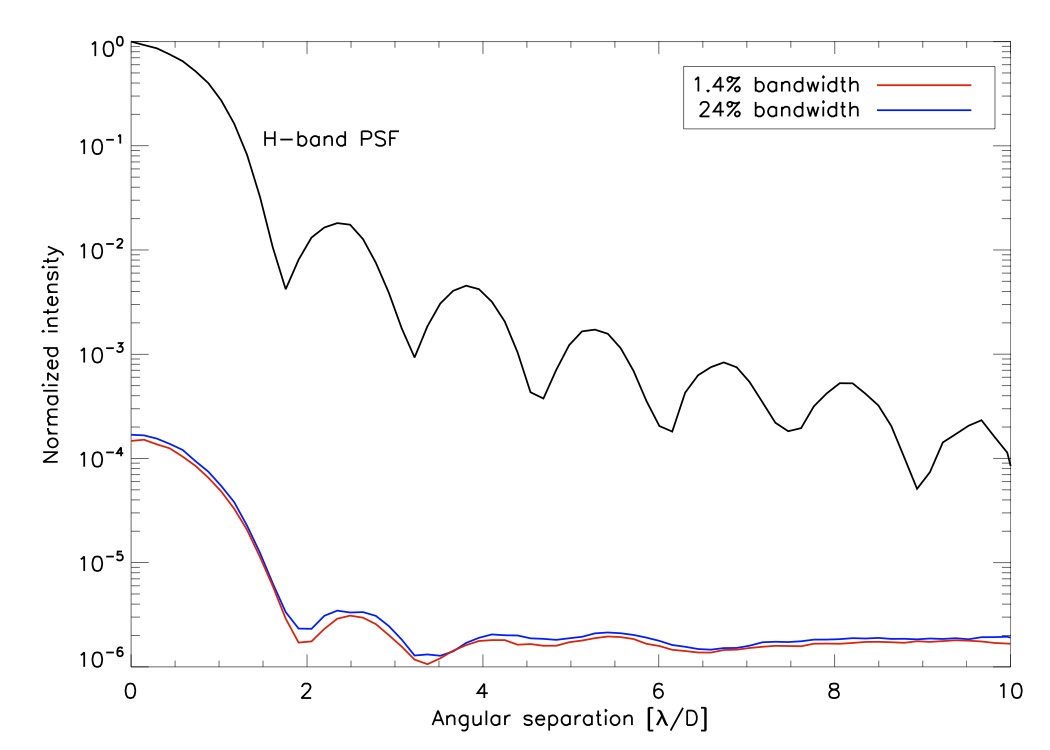}
\includegraphics[width=8.2cm]{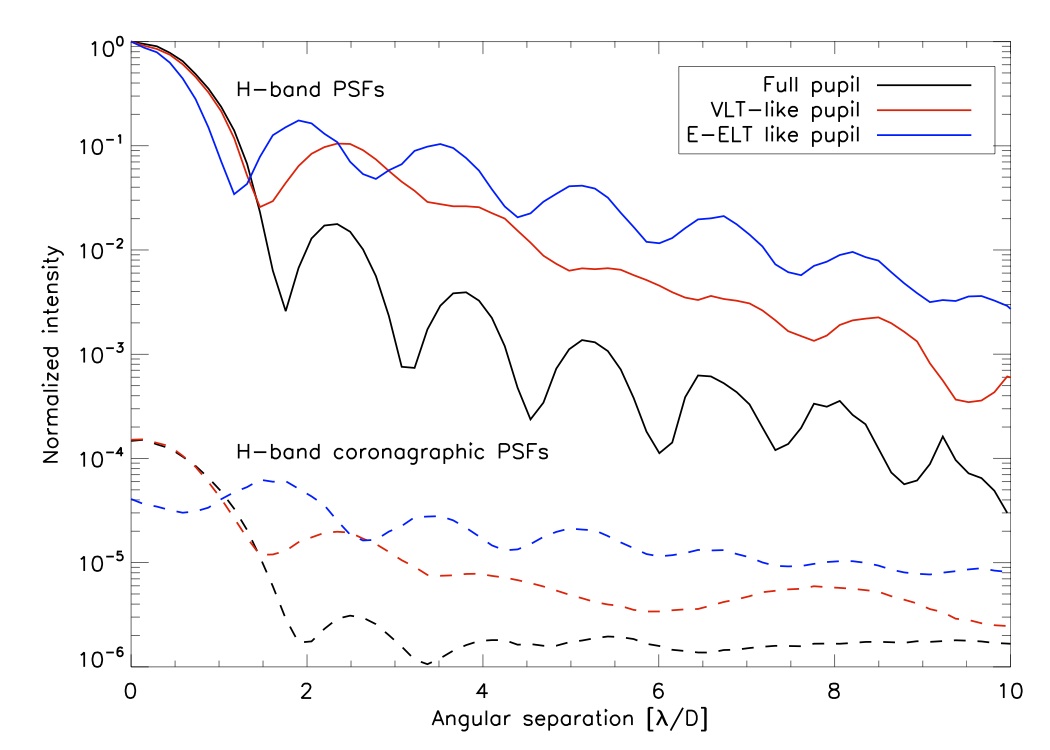}
\end{center}
\caption{Left: experimental results obtained with BL5 with a narrow and broadband filter in the near-IR. Right: experimental results obtained with BL5 with a clear pupil, the VLT pupil, and the E-ELT pupil in the near-IR. Profiles are azimuthally averaged.} 
\label{res1}
\end{figure*}  
\begin{table*}
\centering
\begin{tabular}{llllll}
\hline 
\hline
 & $\tau$ & $\tau_{0}$ & $\mathscr{C}_{5 \lambda/D}$ & $\mathscr{E}$ ($\%$) \\ 
\hline 
Proto 1 & 2410 & 2554  & $3.7\times10^{-5}$ & $\sim 5$  \\
Proto 2 & 5360 & 6792  & $1.7\times10^{-6}$   & $< 1$ \\ 
\hline
\end{tabular}
\caption{Summary of Coronagraphic Results Obtained with BL5: Proto 1 Compared to Proto 2 ($\Delta\lambda/\lambda=24\%$). $\mathscr{E}$ stands for the profile local accuracy.}
\label{resum}
\end{table*}
\begin{figure*}[!ht]
\begin{center}
\centering
\includegraphics[width=4.5cm]{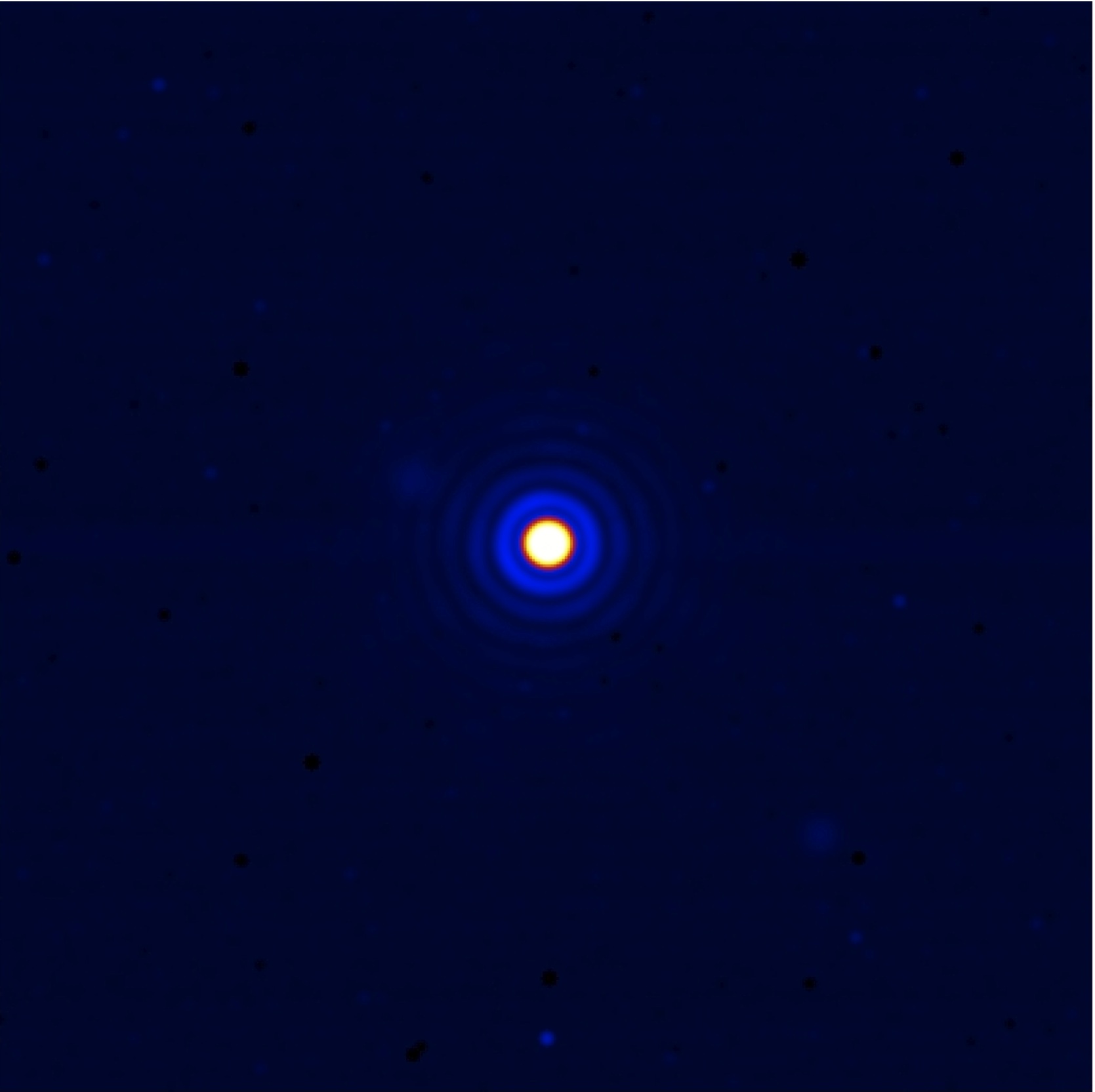}
\includegraphics[width=4.5cm]{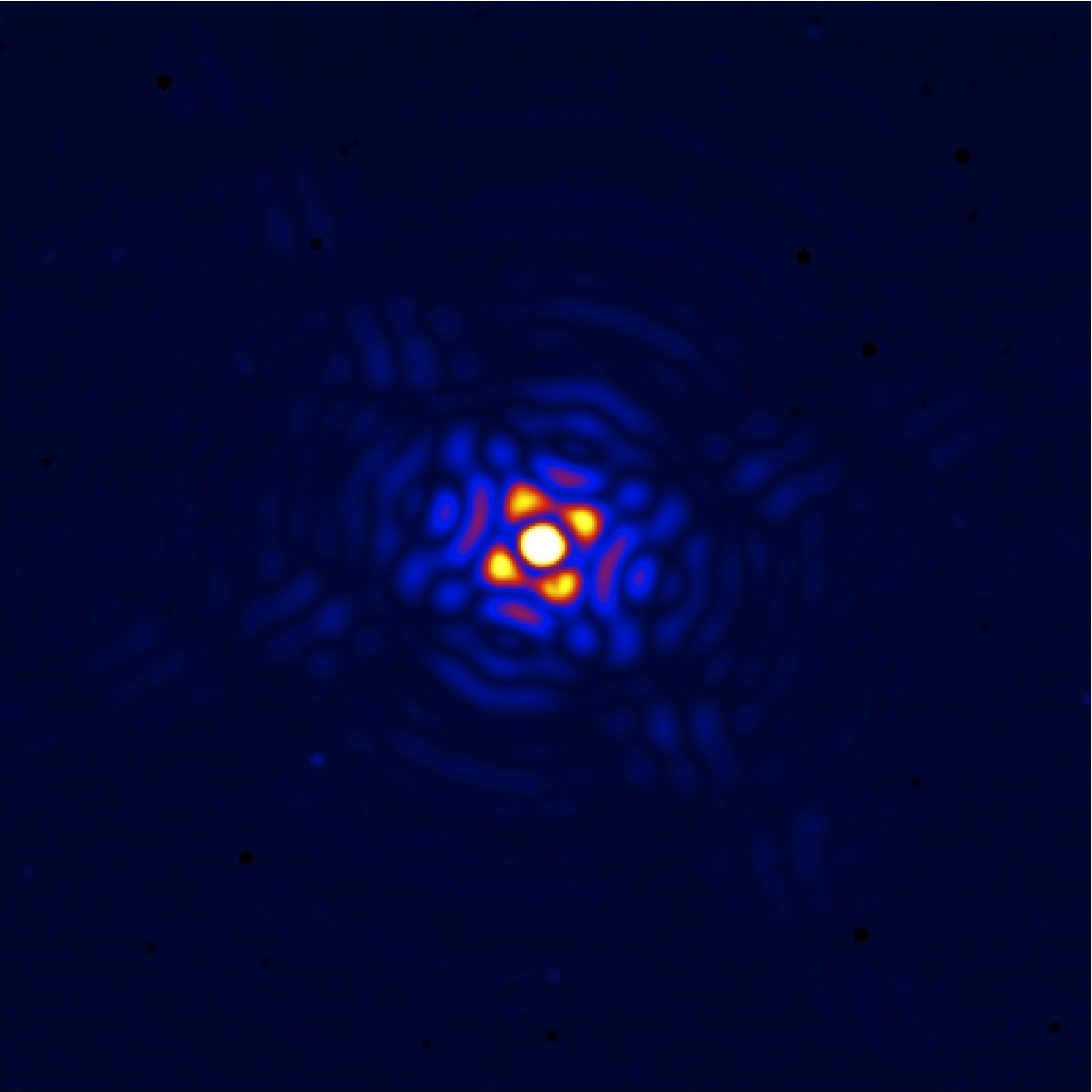}
\includegraphics[width=4.5cm]{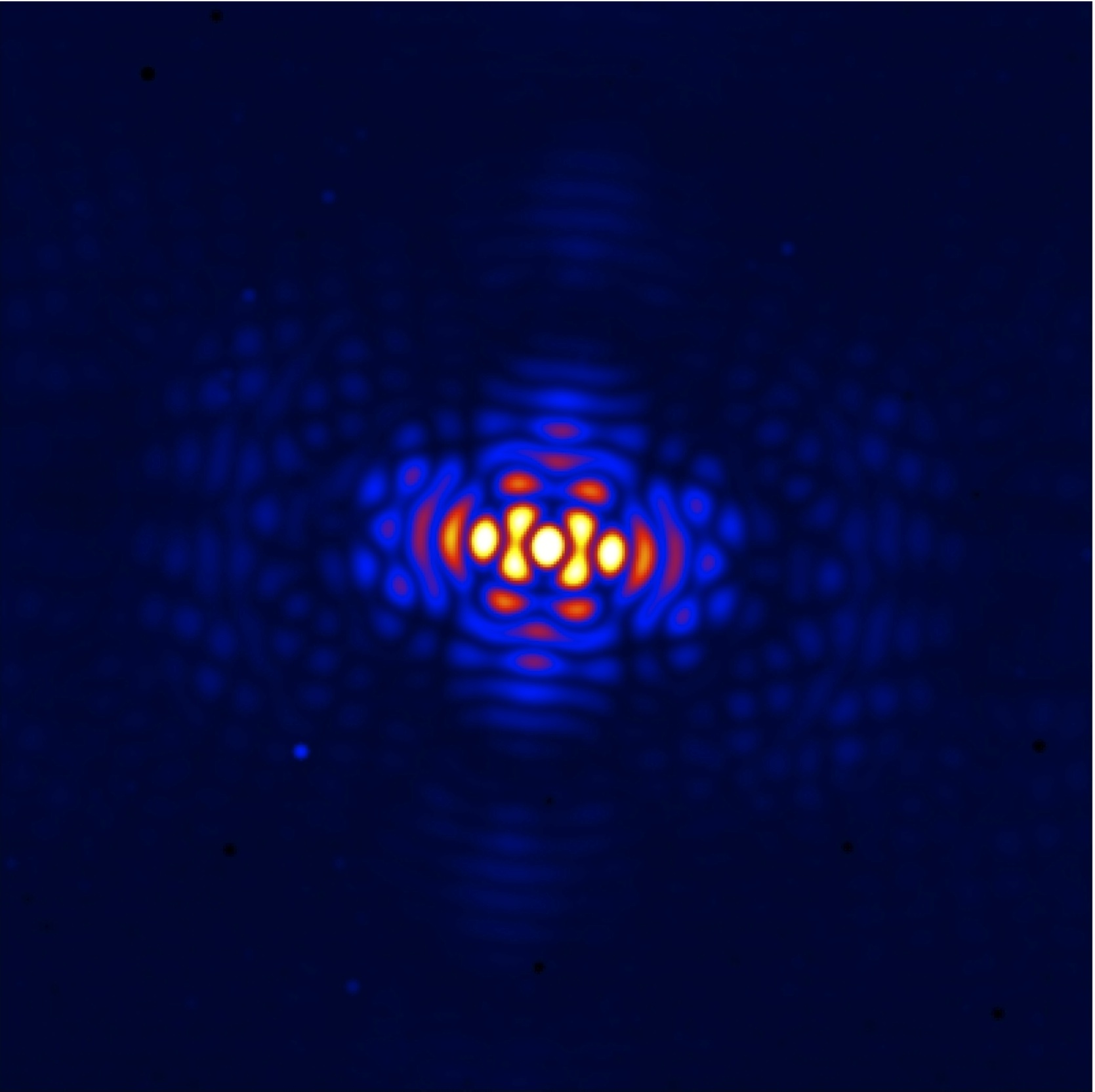}
\includegraphics[width=4.5cm]{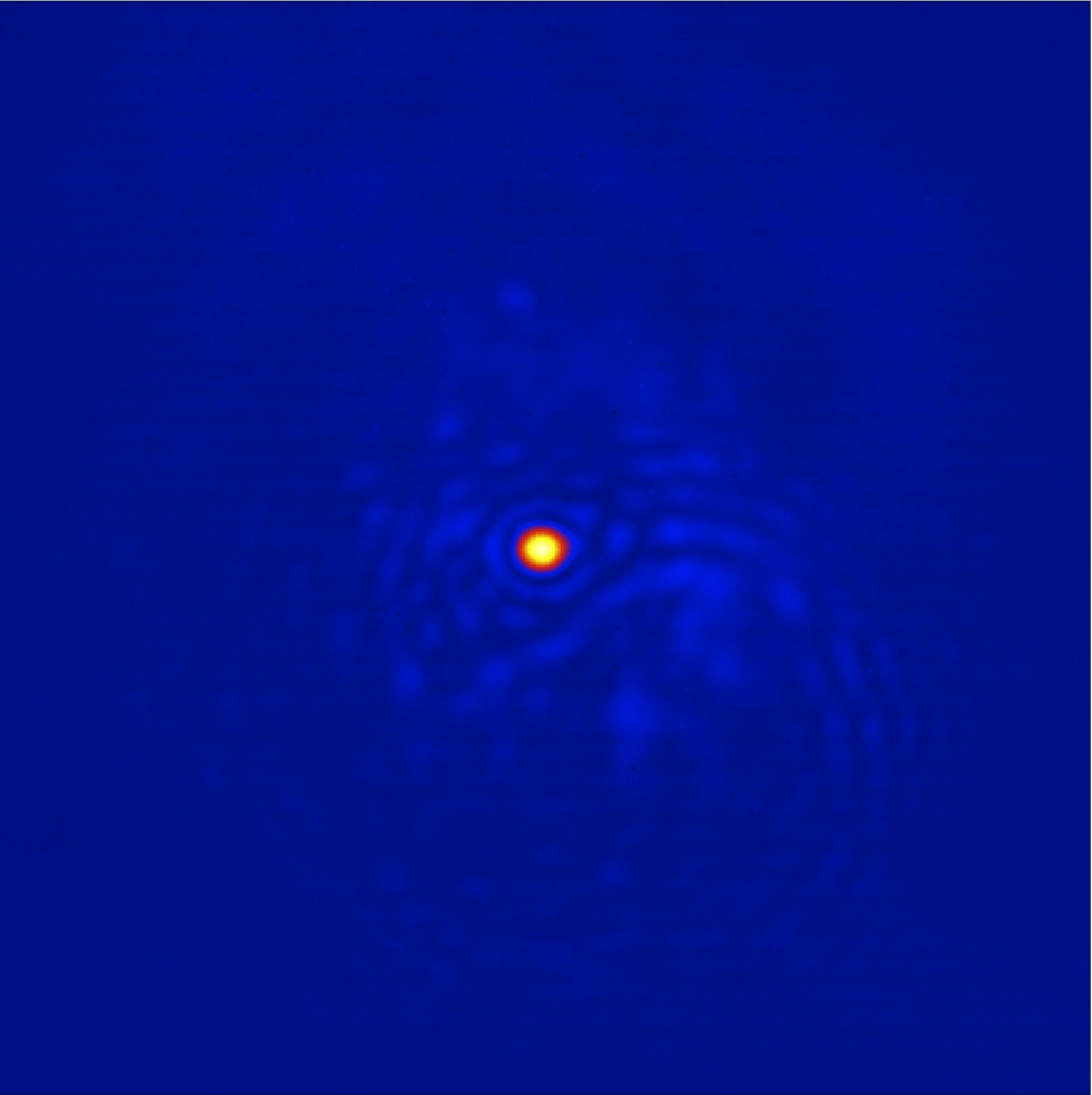}
\includegraphics[width=4.5cm]{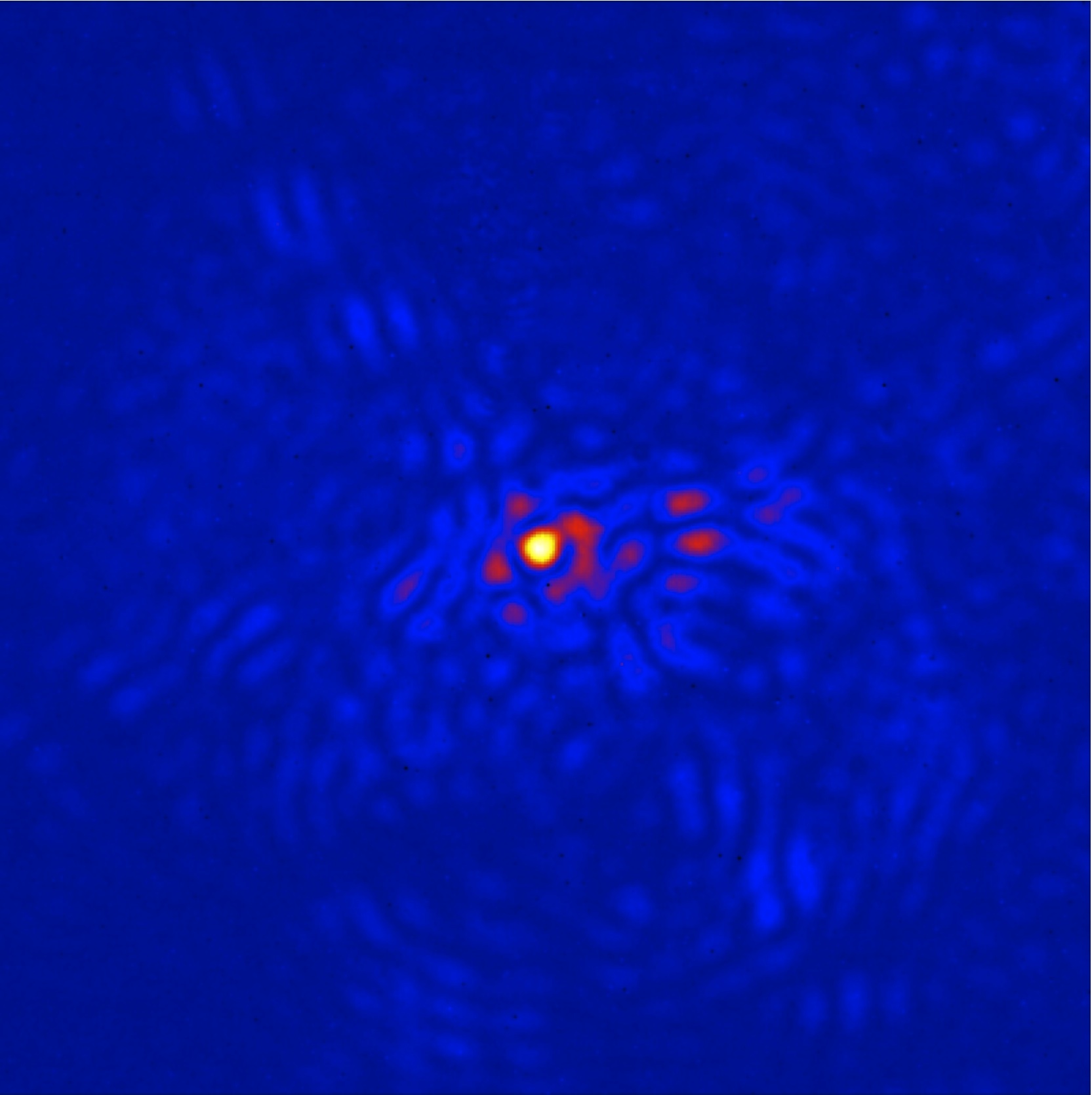}
\includegraphics[width=4.5cm]{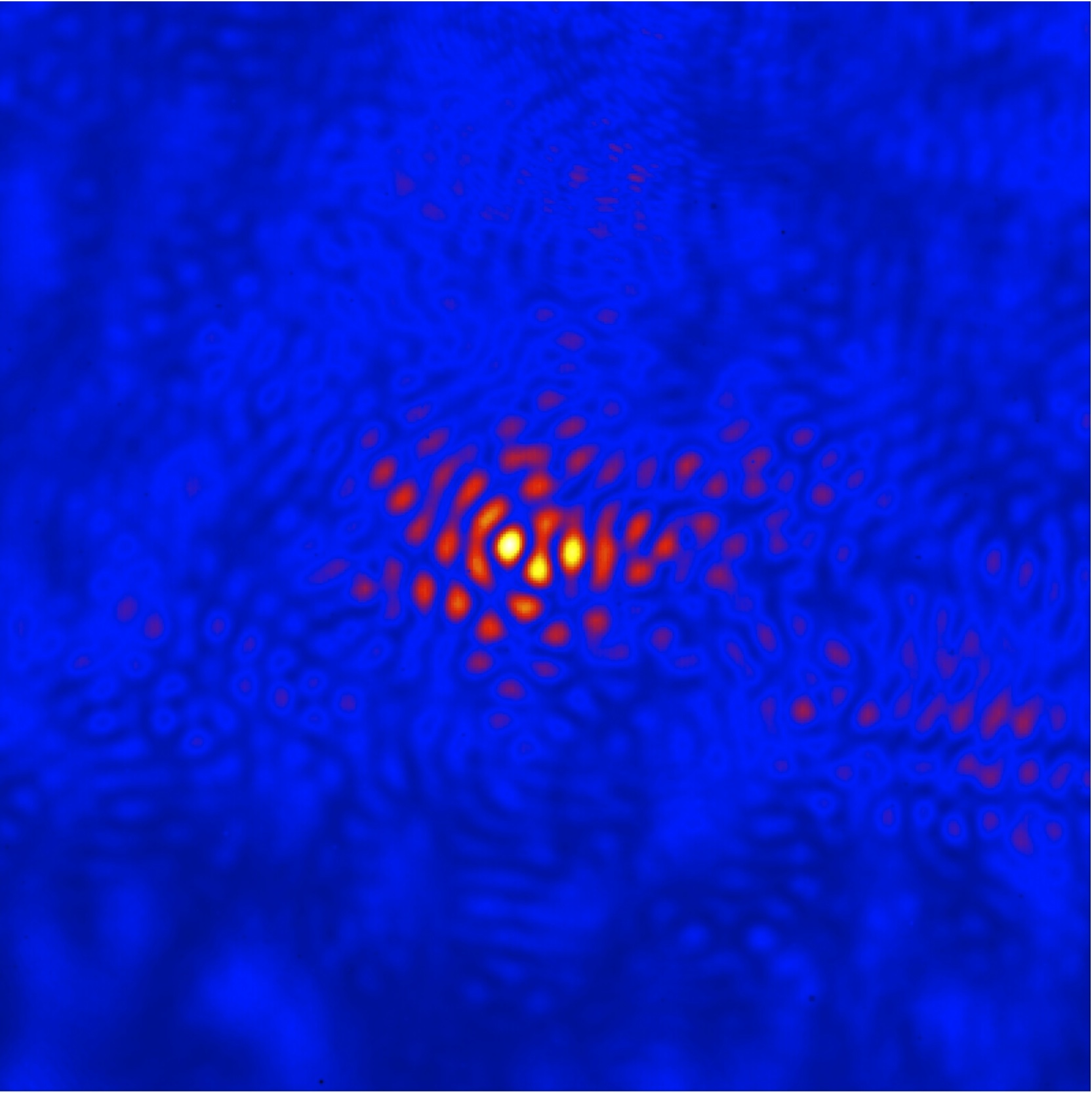}
\end{center}
\caption{PSF (top row) and coronagraphic images (bottom row) obtained with a full circular pupil (left column), the VLT pupil (middle column), and the E-ELT pupil (right column) in the $H$-band. The arbitrary false color distribution has been chosen to enhance contrast for the sake of clarity. } 
\label{PSF}
\end{figure*}  
\subsection{Performance Assessment}
\label{perf}
\subsubsection{Fundamental Performance of Proto 2}
To assess the fundamental performance of Proto 2, we first tested the device with a full circular pupil (Figure \ref{pupil}, top row, left image). 
For the tests described in this sub-section, the experimental conditions are exactly similar to the ones occurring during the evaluation of Proto 1 and reported in Paper 1. The pupil-stop is the same (diameter $D_{stop} = 0.84 \times \Phi$, see FigureÊ\ref{pupil} bottom row, left image), and data have been reduced likewise. Contrast profiles are azimuthally averaged, and were obtained without any active wavefront correction, nor particular data reduction post-processing. Neutral density filters were applied on non-coronagraphic images only. 
Dark frames were obtained by switching off the artificial star source, and the data reduction only corrects bad pixels, background, and scales images by the exposure time and optical density.

Coronagraphic profiles of Proto 2 in the two spectral conditions ($\Delta\lambda/\lambda = 1.4\%$, and 24$\%$) are gathered in Figure \ref{res1} (left), and roughly yields identical performance.
That demonstrates the capability of Proto 2  to achromatically reduce starlight over a wide spectral range.
The contrast evolves from 2$\times10^{-6}$ to $10^{-6}$ from 3$\lambda/D$ to farther angular distances limited by the speckle noise level on the bench.
At IWA, the contrast is 1.7$\times10^{-6}$, which corresponds to an improvement of contrast of a factor of 30  over Proto 1. The peak rejection rate (denoted $\tau_{0}$, being the ratio of the maximum of the direct image to that of the coronagraphic image) is 6792, which is $\sim$2.7 times better than the peak rejection delivered by Proto 1. The total rejection factor (the ratio of total intensity of the direct image to that of the coronagraphic image, denoted $\tau$) follows the improvement likewise and is reported in Table \ref{resum}. The point-spread function (PSF) and coronagraphic images are presented in FigureÊ\ref{PSF} (left column).

While the improvement in profile accuracy ($\mathscr{E}$) over Proto 1 did enhance the peak rejection and contrast level, discrepancy still remains between our actual measurements and the simulation predictions (as formerly reported for Proto 1 in Paper 1). 
At this level of contrast it is difficult to ascertain source of errors, although the scatter-light due to imperfect optical components that create speckles in the scientific image and mask profile errors are first to come to mind. 
Proto 1 was mainly affected by the profile error, which was found to be consistent with the discrepancy found at small angular separations (i.e. peak level error), while the wavefront errors (wavefront high-frequency components) were identified as the dominant source of error in the halo. An overall approximated total amount of wavefront error of  $\sim \lambda/67$ rms at 1.6 microns, (i.e. 24 nm rms) was estimated prior to the pupil stop (Paper 1).
Simulations using the actual profile of Proto 2 shows that the profile error does not preclude to reach deeper contrast than $10^{-6}$, 
while the estimated wavefront error limits the contrast to $\sim 10^{-7}$. Thus, it does not explain our current limitation of contrast. 
However, our bench is not in a clean room environment and dust contamination is significant on all our optics. Given the small size of the pupil, a dust density map estimation over the pupil area has to be taken into account for proper simulation analysis. Because the speckle noise level on the bench is at the level of $10^{-6}$ and cannot be explained by manufacturing aspects of Proto 2, or wavefront error estimation of the optical components of our setup, it is believed that dust is the main contributor to our current limitation.

\subsubsection{Comparison with Other Mature Developments}
A first order comparison between our research efforts in the implementation of a BLC in microdots and the one presented in \citet{Krist09} can be initiated to further validate the maturity and validity of the approach.
Although,  functions, bandwidths, dots distribution algorithms, the way fabrication errors are accounted for, and the testbed are not exactly similar, the fundamental parameters are comparable.
The Bessel-squared radial profile BLC developed by \citet{Krist09}  has a 6$\lambda/D$ IWA (1$\lambda/D$ larger than our BL5 prototype), and a $s$ parameter equal to 19 ($F_{\#}$=18, $\lambda$=2.2$\mu$m, $p$=2.1$\mu$m), whereas $s=16$ for our prototype. The radial profile error is not explicitly discussed in \citet{Krist09}, nonetheless from Figure 7 of their paper, the local error seems to be at the level of few $\%$, being mostly localized in the inner part of the prototype. Their qualitative comparison between non-coronagraphic and coronagraphic images suggests that the peak rejection $\tau_0$ is close to $10^{4}$, which is comparable to the performance presented in this paper with our BL5. 
Both developments therefore lead to similar performance, being one order of magnitude below the level of the theoretical expectations presented in Paper 1 (Figure 5), where we show with numerical simulations that the peak rejection limitation due to the presence of the dots is $10^{5}$ when $s$ and the IWA are in the order of values considered here. 
In both developments, it is believed that speckles due to aberrations in the testbed are the origin of the contrast limit.

\subsection{Application to Arbitrary Telescope Apertures}
\subsubsection{Complex Pupils}
Because this study was originally realized in the framework of the future E-ELT high-contrast instrument \citep[EPICS,][]{EPICS}, we tested Proto 2 with the telescope pupil planned for the E-ELT, though we do not consider here its segmentation aspect. In addition,  since actual telescopes can benefit from BLC capabilities, we extended our test of Proto 2 to the case of the VLT pupil. Both aperture designs (VLT and E-ELT) are shown in Figure \ref{pupil} (top row, middle, and right images).
Our pupil masks are made in a laser-cut, stainless-steel sheet to an accuracy of 0.002 mm. The pupil diameter for all masks remains similar as in Section \ref{perf} ($\Phi$=3mm).
The VLT pupil is designed with the central obscuration scaled to 0.47 mm $\pm$ 0.002 mm (14$\%$ linear ratio) and the spider-vane thickness is 15$\mu$m $\pm$ 4$\mu$m. For the E-ELT pupil, the central obscuration ratio is 29$\%$ (0.88 mm $\pm$ 0.002 mm), and the spider-vane thickness is 40$\mu$m $\pm$ 4$\mu$m. 
The corresponding pupil stops used in the experiment are shown in Figure \ref{pupil} (bottom row, middle and right images). The BLC pupil-stop throughput for the VLT aperture is 42$\%$, while the one for the E-ELT pupil is $\sim20\%$. The optimization of the BLC pupil stops emphasized the achromaticity aspect (24$\%$ spectral bandwidth) and allowing the deepest contrast possible considering the achievable contrast imposed by the presence of the dots in the design ($10^{-7}$ to $10^{-8}$ contrast at 5$\lambda/D$, see Paper 1). Due to the complexity of the E-ELT aperture, it imposes a severely reduced pupil stop, but one might note that its optimization can be relaxed to accommodate less ambitious contrast goals (e.g., $10^{-6}$). This will be further analyzed later for the case of the VLT pupil.

PSF and coronagraphic images recorded during the experiment are presented in FigureÊ\ref{PSF} (middle and right columns). 
Contrast profiles are shown in Figure \ref{res1} (right), which gathered performance obtained with all pupil configurations (full circular pupil, VLT-pupil, and E-ELT pupil). Due to the complexity of the VLT and E-ELT pupils, their corresponding PSFs exhibit various diffraction patterns (central core and spider diffraction spikes), and thus different PSF contrast levels (the power-law of the PSF intensity as a function of the angular separation deviates from the pure full pupil configuration). Whereas coronagraphic contrasts degrade with the complexity of the pupil, the gain with the PSF level is roughly conservative (Figure \ref{res1}). The BLC delivers 1.7$\times10^{-6}$,  5$\times10^{-6}$, and 2$\times10^{-5}$ contrast levels at 5$\lambda/D$ angular separation over 24$\%$ bandwidth, with a full pupil, the VLT pupil, and the E-ELT pupil respectively. 
This indicates that the BLC is able to maintain high-contrast levels with large central obscuration and spiders, unlike several coronagraphic concepts.  
\begin{figure}[!ht]
\begin{center}
\includegraphics[width=6.0cm]{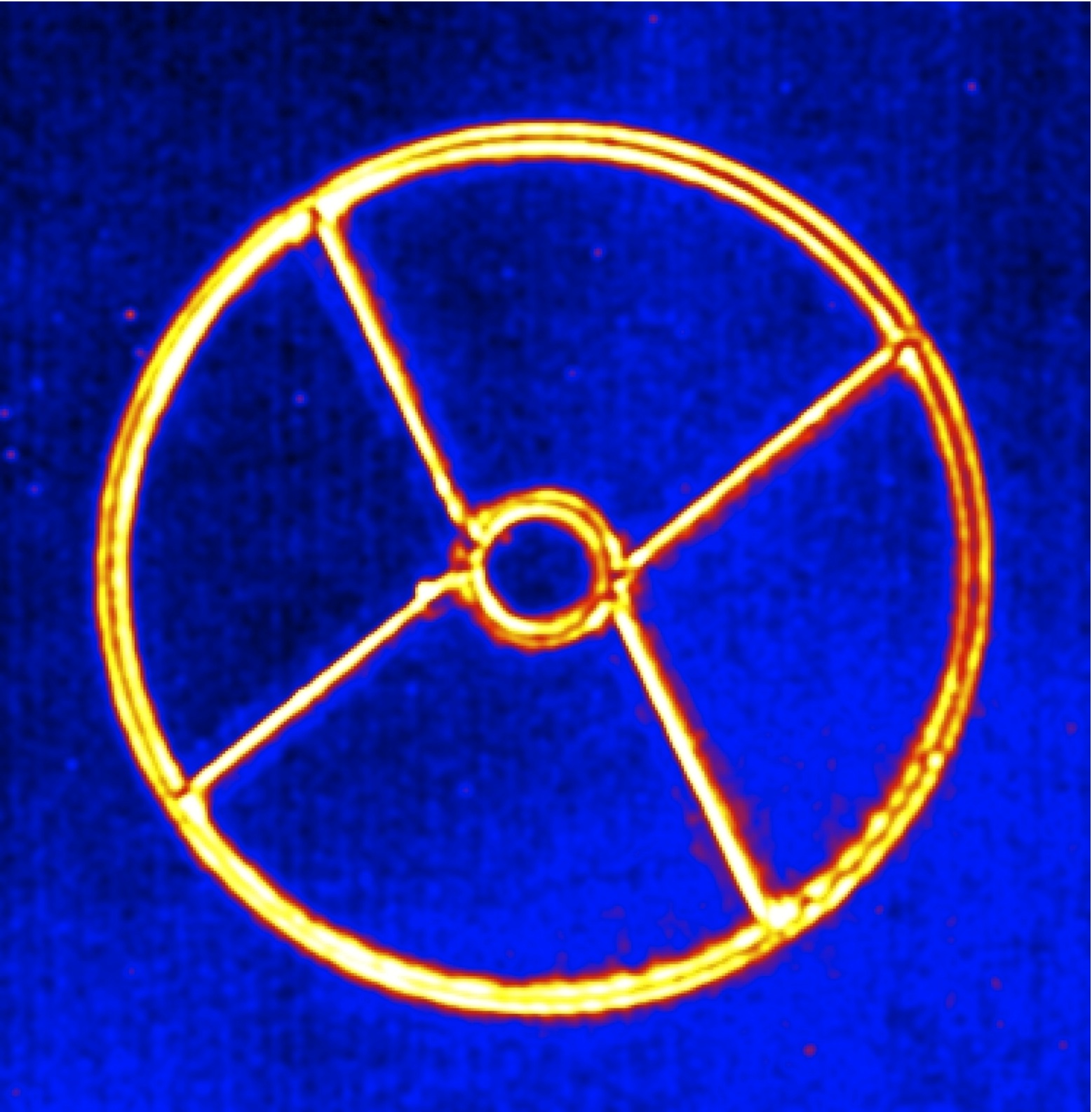}
\end{center}
\caption{Coronagraphic pupil image before the high-spatial frequency filtering of the pupil stop recorded in $H$-band, while using the VLT pupil as the entrance aperture. The image shows the distribution of the diffracted light that is strictly localized in the vicinity of the pupil edges and spider arms. The arbitrary false color distribution has been chosen to enhance contrast for the sake of clarity. } 
\label{pupilBL5}
\end{figure}  
\begin{figure*}[!ht]
\includegraphics[width=4.cm]{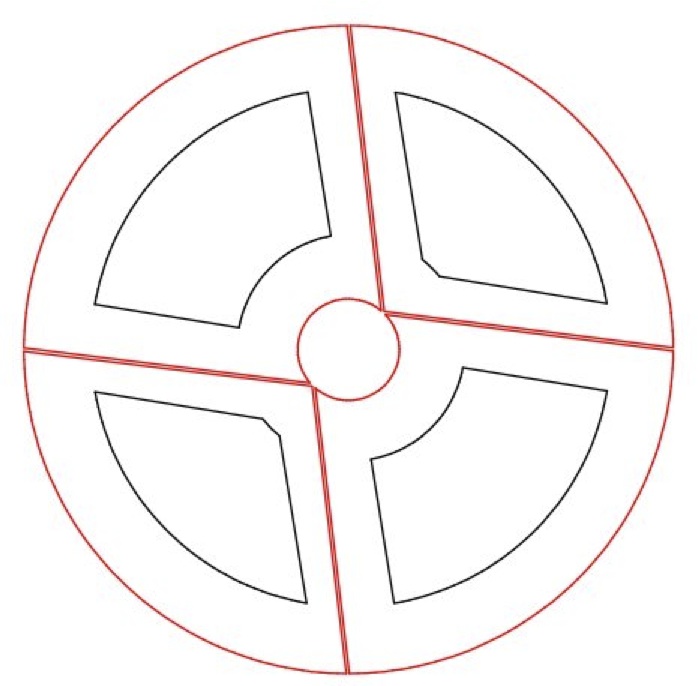}
\includegraphics[width=4.cm]{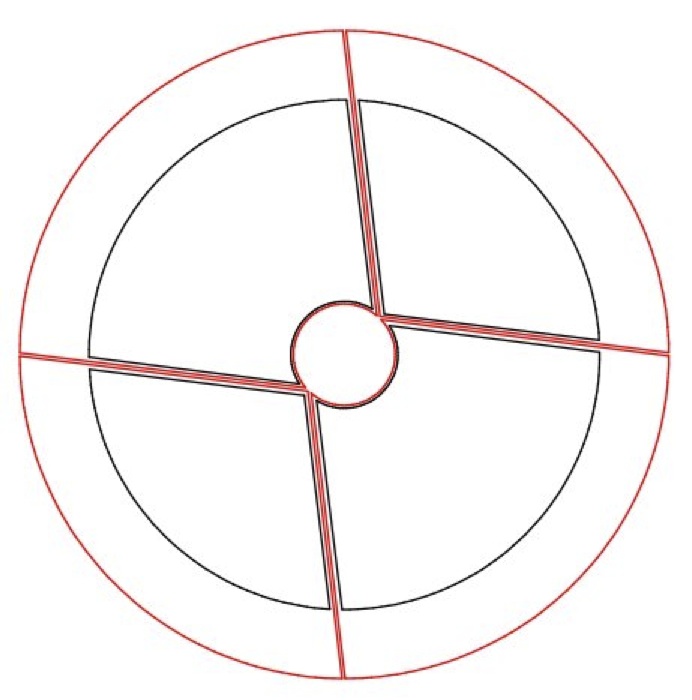}
\includegraphics[width=4.cm]{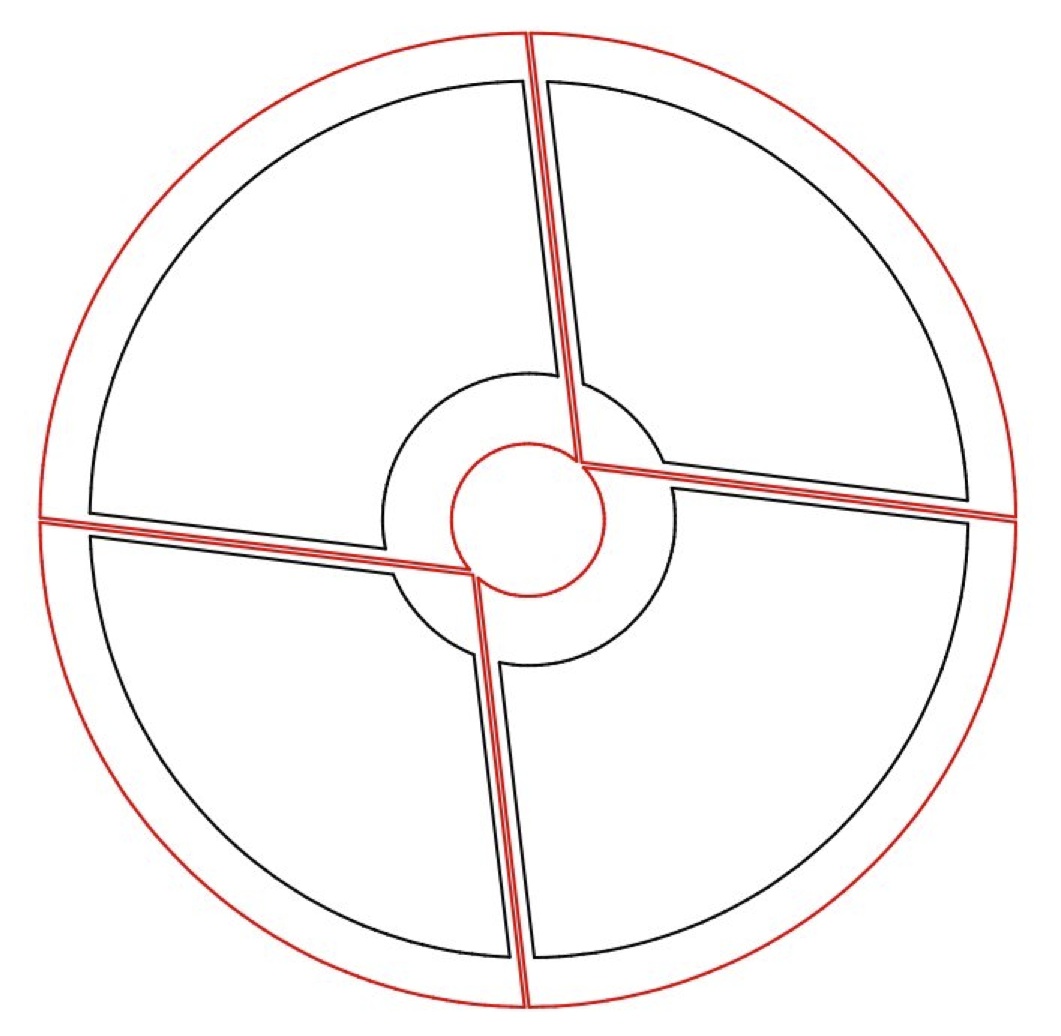}
\includegraphics[width=4.cm]{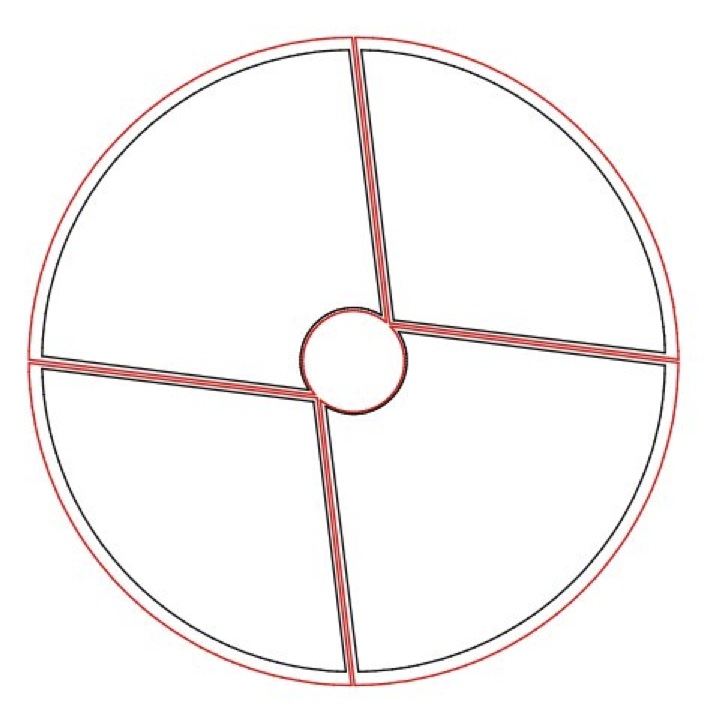}
\caption{From the left to the right: specified pupil stops (black) superimposed to the VLT-pupil (red) corresponding to 42$\%$, 60$\%$, 70$\%$, and 92$\%$ throughput.}
\label{STOPS}
\end{figure*} 
\subsubsection{Pupil-Stop Discussion}
\begin{figure}[!ht]
\begin{center}
\includegraphics[width=9.0cm]{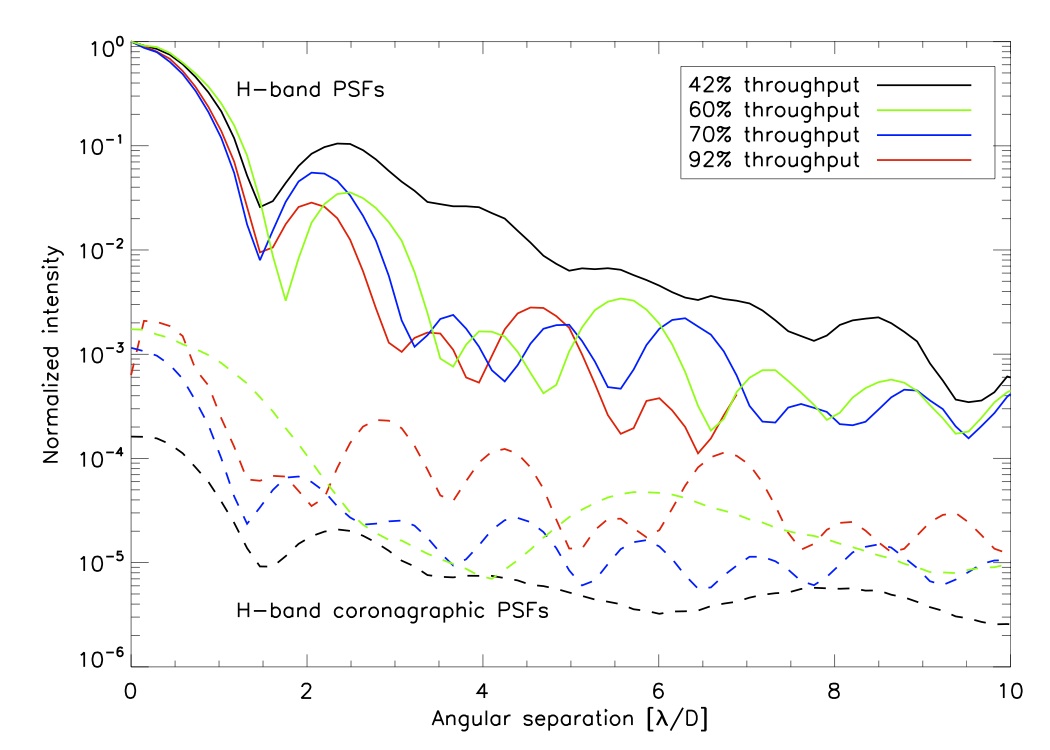}
\end{center}
\caption{$H$-band experimental contrast levels reached with the BLC with various pupil-stops (VLT configuration).} 
\label{res3}
\end{figure}

Figure \ref{pupilBL5} presents the coronagraphic pupil image before the filtering by the pupil-stop. In this case the entrance aperture is the VLT mask.
It is clearly observable that the diffracted light is strictly localized in the vicinity of the pupil edges and spider arms; thus, filtering these areas is mandatory, whereas leaving some room in the design of the pupil-stop. To assess the impact of the pupil-stop optimization, we compared the performance delivered by Proto 2 with its initial pupil-stop to that of various alternative designs. The set of pupil-stops used in the experiment is shown in Figure \ref{STOPS}, superimposed on the VLT pupil for the sake of clarity. From the left to the right are presented: (1) the initial pupil-stop (used in Section \ref{perf}), (2) a 60$\%$ throughput pupil-stop minimizing the impact of the diffracted light on the pupil edges, (3) a 70$\%$ throughput pupil-stop minimizing both areas on the pupil edges and around the central obscuration, (4) a 92$\%$ throughput pupil-stop with little shrinking of the central obscuration and pupil diameter. Spider arms are treated almost alike by all the masks. 
Results are presented in Figure \ref{res3}.  All configurations provide coronagraphic gain, with better contrast  when both pupil diameter and central obscuration are addressed through the optimization (black and blue curves). In a wide spectral bandwidth regime ($24\%$), a contrast from $10^{-5}$ down to $10^{-6}$ at  angular separations farther than 5$\lambda/D$ is reachable with a pupil stop throughput that can basically stand in the range 40 to $70\%$. This still provides an excellent stellar peak rejection.  This result confirms that the BLC can deliver high-contrast levels with system transmission comparable to other coronagraphic concepts in the face of complex pupil apertures.

\section{Conclusion}
We described the development and laboratory experiment of a second-generation BLC mask using a microdot design in the near-infrared.
Thanks to the improvement of the calibration of the manufacturing process through multiple variations of the microdot spatially distribution design, the impact of fabrication errors have been reduced. 
The local accuracy of the BL5 profile was reduced to below 1$\%$. This basically corresponds to a factor of five improvement over the initial prototype, whereas contrast levels have been enhanced by more than an order of magnitude, while the peak rejection rate is doubled.
This advanced $H$-band band-limited device demonstrated excellent contrast levels down to $\sim$ $10^{-6}$ at angular separations farther than 3$\lambda/D$ over 24$\%$ spectral bandwidth, being limited by the speckle noise on the optical bench.  In this context, the halftone-dot technology is confirmed as a mature solution for the development of a BLC. 

Current and next generation high-contrast instruments can directly benefits from such performance, as more emphasis is placed on associating high-contrast suppression systems and speckle calibrations over a large spectral band. Additionally, since wavefront errors downstream of the coronagraph produce quasi-static speckles with intensities that are proportional to the residual coronagraphic peak intensity, the use of such device that allows deep peak rejection arise as a major advantage.
With these considerations in mind, we experimentally explored the capabilities of the BLC to maintain high contrasts with complex telescope apertures. We show that 
in a wide spectral bandwidth regime ($24\%$), a contrast from $10^{-5}$ down to $10^{-6}$ at angular separations farther than IWA is possible with a high stellar peak rejection ($>1000$), when the BLC is combined with a high throughput pupil-stop ($70\%$).
        
The activity outlined in the paper was supported by the European Commission, Seventh Framework Programme (FP7), Capacities Specific Programme, Research Infrastructures; specifically the FP7, Preparing for the construction of the European Extremely Large Telescope Grant Agreement, Contract number INFRA-2007-2.2.1.28.
P.M. acknowledges Precision Optical Imaging and Aktiwave (Rochester, New York) for the free delivery of this improved BL5 prototype. 

\nocite{*}

\bibliography{MyBiblio2}
\end{document}